\documentclass[runningheads]{llncs}

\usepackage{amsmath}
\usepackage{multirow}
\usepackage{array}
\usepackage{booktabs}
\usepackage{graphicx}
\usepackage{subcaption}
\usepackage{float}

\usepackage{adjustbox}
\usepackage[table,xcdraw]{xcolor}

\usepackage[table,xcdraw]{xcolor}
\usepackage[table,xcdraw]{xcolor}
\begin{document}
\title{ArPA: A Novel Speech Analysis and Correction Tool for Arabic-Speaking Children}
\titlerunning{ArPA: A Novel Speech Analysis and Correction Tool for Arabic-Speaking Children}
%

\author {Lamia Berriche \inst{1} \orcidID{0000-0002-5413-1335},
Maha Driss\inst{2,3}\orcidID{0000-0001-8236-8746}, Areej Ahmed Almuntashri \inst{4}, Asma Mufreh Lghabi \inst{4}, Heba Saleh Almudhi \inst{4}, Munerah Abdul-Aziz Almansour \inst{4}} 
%
\institute{ CS Department, CCIS, Prince Sultan University, Riyadh 12435, Saudi Arabia\and RIOTU Lab, CCIS, Prince Sultan University, Riyadh 12435, Saudi Arabia \and RIADI Laboratory, University of Manouba, Manouba 2010, Tunisia \and Imam Mohammad Ibn Saud Islamic University, Riyadh, Saudi Arabia
 \let\thefootnote\relax\footnotetext{This work is supported by Prince
Sultan University in Saudi Arabia}
}
\maketitle              
\begin{abstract}
This paper introduces a new application named ArPA for Arabic kids who have trouble with pronunciation. Our application comprises two key components: the diagnostic module and the therapeutic module. The diagnostic process involves capturing the child's speech signal, preprocessing, and analyzing it using different machine learning classifiers like K-Nearest Neighbors (KNN), Support Vector Machine (SVM), and Decision Trees as well as deep neural network classifiers like ResNet18. The therapeutic module offers eye-catching gamified interfaces in which each correctly spoken letter earns a higher avatar level, providing positive reinforcement for the child's pronunciation improvement. Two datasets were used for experimental evaluation: one from a childcare centre and the other including Arabic alphabet pronunciation recordings. Our work uses a novel technique for speech recognition using Melspectrogram and MFCC images. The results show that the ResNet18 classifier on speech-to-image converted data effectively identifies mispronunciations in Arabic speech with an accuracy of 99.015\% with Mel-Spectrogram images outperforming ResNet18 with MFCC images. 
\end{abstract}
\section{Introduction}
Speech disorders severely impact the child's communication abilities and academic progress. The impact of speech disorders on children's social development and educational achievement is considerable \cite{cowan2020language}. Among these disorders are difficulties with pronunciation, rhythm, and voice quality. These disorders can have a variety of negative repercussions, such as reduced self-esteem, dissatisfaction, shame, and, in severe cases, withdrawal from social engagement \cite{lloyd2020peer}.
It is important to be aware that errors in pronunciation are an inevitable part of language development in children, with research revealing that a substantial number of children have such difficulties during early speech acquisition \cite{cowan2020language}. Because of the linguistic complexities of Arabic, Arabic-speaking children have specific pronunciation disorders. The complexity of Arabic phonology, as well as the differences between Arabic and other languages, are the main causes of these disorders. While many pronunciation difficulties may resolve spontaneously as children acquire their language abilities, certain specific difficulties may remain and need tailored assistance. Early intervention plays an important role in developing effective communication skills and language abilities in Arabic-speaking children.
In this context, this paper proposes an innovative application named \textbf{ArPA} (\textbf{Ar}abic \textbf{P}ronunciation \textbf{A}id) developed particularly for Arabic-speaking children who struggle with pronunciation. This application is composed of two main modules: 1) diagnostics and 2) therapeutic interventions. The diagnostic module entails recording and evaluating the child's voice signal using a variety of machine and deep learning classifiers. The therapeutic intervention module uses gamified interfaces that support correct pronunciation. Two datasets were used to test the performance of several classifiers, with the ResNet18 classifier proving effective in detecting mispronunciations. 
The main contributions of this paper are summarized in the following points:
\begin{itemize}
\item Collection of an audio dataset for the Arabic alphabet featuring children's correct and mispronounced pronunciations. Various pronunciation variations can be captured in this dataset to ensure that models are trained reliably.
\item Investigation of traditional machine learning algorithms for pronunciation classification, focusing on methods such as Decision Trees, K-Nearest Neighbors (KNN), and Support Vector Machine (SVM) to establish baseline performance.
\item Exploration of image-pretrained deep learning models adapted for audio data, employing techniques such as transfer learning. 
\item Conduction of a comprehensive evaluation of the performance of all models using multiple metrics, including accuracy, precision, recall, and F1-score. T
\item Development of an innovative application named \textbf{ArPA} for Arabic children with pronunciation difficulties. 
\end{itemize}
The structure of this paper is organized as follows: Section \textbf{2} provides a comprehensive literature review about pronunciation classification with a focus on Arabic pronunciation applications. Section \textbf{3} outlines the architecture of the proposed application \textbf{ArPA} and the techniques employed in our study. Section \textbf{4} discusses the datasets used, the experiments conducted, and the results obtained, providing a thorough analysis of the performance of various machine learning and deep learning models. Finally, Section \textbf{5} concludes the paper.
\section{Related Work}
Pronunciation classification has made significant advances over the past few years \cite{lounis2024mispronunciation}. Nowadays, automated systems for diagnosing pronunciation errors are crucial for assisting learners in enhancing their speaking abilities. The purpose of this section is to examine the current literature about approaches used for assessing pronunciation, specifically emphasizing strategies that address the distinct phonetic difficulties encountered by Arabic learners.

The study presented in \cite{nazir2019mispronunciation} investigated the use of deep CNNs in educational systems for identifying incorrect pronunciations of Arabic phonemes. The authors suggested two methods: utilizing CNN features and employing transfer learning. Utilizing features extracted from CNN layers, KNN, SVM, and NN classifiers were trained. The CNN model based on transfer learning performed better than a baseline that used manually crafted acoustic-phonetic features, achieving higher accuracy than state-of-the-art methods.

In \cite{akhtar2020improving}, authors proposed a CNN model based on AlexNet to analyze spectrograms of Quranic verses and detect mispronounced Arabic words. The proposed model outperformed transfer learning and hand-crafted feature methods.

To address the minor pronunciation variations in Arabic speech processing, the study presented in \cite{asif2021approach} proposed a classification system for differentiating between correct and erroneous pronunciations of Arabic short vowels. To guarantee adequate data volume, a new dataset of recorded Arabic alphabet sounds, which included every possible vowel pronunciation, was created for this study. Using this dataset, a deep CNN was trained, and it was able to classify audio into 84 distinct phoneme classes with high accuracy.

In \cite{algabri2022mispronunciation},  Algabri et al. proposed a pronunciation training system using deep learning for non-native Arabic speakers that can identify mispronunciations and give feedback on articulation. The system surpassed current methods in phoneme recognition, mispronunciation detection, and identifying articulatory features by analyzing spectral images and recognizing multiple labels. Neural text-to-speech technology was used to create a dataset of typical substitution errors.

Using deep learning techniques, particularly LSTM networks, the study presented in \cite{ahmed2023arabic} introduced a two-level detection framework to identify mispronunciations and speaker gender in Arabic. The hyperparameters were optimized through grid search using a two-stage approach, setting a baseline for mispronunciation detection. The LSTM model performed better than the latest methods, greatly increasing accuracy in identifying and detecting gender.

Lounis et al. in \cite{lounis2024anomaly} proposed a novel method for Arabic mispronunciation detection using a variational autoencoder (VAE) and anomaly detection to address sparsely labeled data. The VAE identified mispronunciations as "abnormal" and accurate pronunciations as "normal." Tested on the ASMDD dataset, the VAE outperformed state-of-the-art CNNs and traditional autoencoders in detecting Arabic mispronunciation.

These studies have significant drawbacks. Primarily, several proposed models focus on particular aspects of pronunciation, such as short vowels or typical substitution errors, consequently omitting other important mispronunciation types. Additionally, the use of anomaly detection in certain approaches may result in misclassifications, and hyperparameter adjustments can be time-consuming, with no guarantees of achieving optimal results. Our application, \textbf{ArPA}, is proposed to overcome the previously detailed drawbacks. It is specifically designed for kids with various pronunciation issues, as solutions geared towards adults may not be suitable for children's specific speech patterns. In contrast to prior studies focusing on adults or non-native Arabic speakers, ArPA caters to children's unique needs using game-based treatment and specialized datasets designed for children. By utilizing deep learning, ArPA attains a high level of precision in detecting mispronunciations, making it more efficient and appropriate for young learners in comparison to current solutions designed for adults.
\section{Proposed Approach}
\begin{figure}
    \centering
    \includegraphics[width=1 \linewidth]{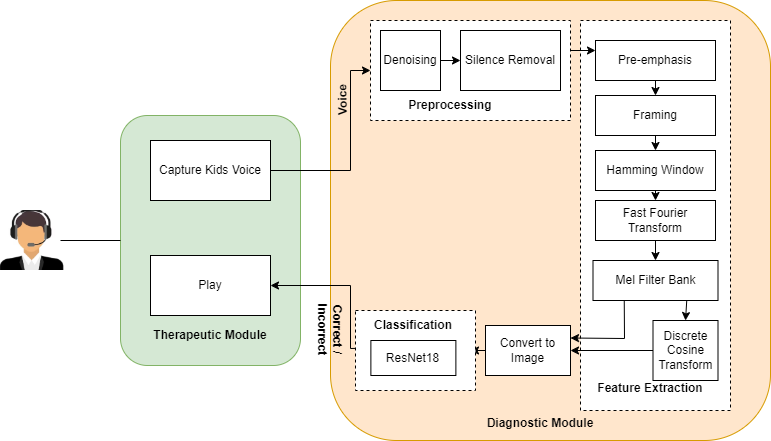}
    \caption{ArPA Architecture}
    \label{fig:ArPA_Architecture}
\end{figure}

Our application aims to evaluate children's pronunciation of Arabic letters and then encourage improvement through an engaging game. \textbf{ArPA} is composed of two modules: the therapeutic module and the diagnostic module, see \ref{fig:ArPA_Architecture}. 
\begin{enumerate}
    \item \textbf{Therapeutic Module:} First, this module is responsible for capturing the kid's voice and delivering it to the diagnostic module. After the kid's pronunciation is evaluated by the diagnostic module, this module proposes a tailored and engaging game for kids. Interactive and enjoyable activities are used to practice incorrectly pronounced letters. The kid's progress is saved in a report which could be viewed by the parents and the therapist. 
    \item \textbf{Diagnostic Module:} The diagnostic module starts with preprocessing the child's raw voice by denoising it and removing the silence. After that, Melspectrogram and Mel-Frequency Cepstral Coefficients (MFCC) features are extracted and converted to images. Finally, a trained ResNet18 classifier evaluates the correctness of the child's pronunciation. 
    \end{enumerate}

\subsection{Preprocessing}
In this phase, the signal undergoes a denoising and a silence removal processes.  The main challenge in noise reduction is mitigating external disturbances while preserving the integrity of the original speech signal. We applied a Gaussian filter which suppresses the high-frequency noise components leading to a smoother signal \cite{das2021fundamentals}.
\subsection{Feature Extraction}
 In this work, we used Mel-Frequency Cepstral Coefficients (MFCC) and Melspectrogram features. Mel-spectrogram is a time-frequency representation that maps the power spectrum of an audio signal to the mel scale whereas MFCCs are obtained after taking the Discrete Cosine Transform (DCT) of the log mel-spectrogram. The cepstral coefficients are computed based on known human auditory perception. To extract MFCC features, we started by applying a first-order high-pass filter to increase the energy level of high frequencies. Then, we segmented the signal into L-sized overlapping frames with size L, where (20ms $<$ L $<$ 40ms). Afterwards, a time domain Hamming window is applied to minimize the discontinuities between frames as shown by Eq.\ref{eq:Windowing1}.
\begin{equation}
\begin{aligned}
y(n) &= s(n) \cdot w(n) \\
\text{with} \quad w(n) &= 0.54 - 0.46 \cos \left( \frac{2\pi n}{N} \right), \quad 0 \leq n \leq N
\end{aligned}
\label{eq:Windowing1}
\end{equation}
where Where $N$ is the number of samples in each frame, $s(n)$ is the input speech signal, $y(n)$ is the windowed output signal and $w(n)$is the Hamming window.
Next, we applied a Fast Fourier transform as represented by Eq.\ref{eq:Fourier} to each windowed frame.
\begin{equation}
S(k) = \sum_{i=0}^{N-1} s(i) e^{j \frac{2 \pi k i}{N}}
 \label{eq:Fourier}
\end{equation}
$S(k)$ is the $k^{\text{th}}$ Fourier transform coefficient and $s(i)$ is the $i^{\text{th}}$ speech signal sample. 
After that, the Mel filter bank is applied to the spectrum of each frame and the Log energy spectrum is computed representing the Mel-Spectrogram. To obtain the MFCC, the log Mel scale spectrum is converted to a time domain using DCT. 
\subsection{Image conversion}
In this phase, we convert the MFCC and Mel-Spectrogram images by scaling the data values in the matrices to the range of a colormap. The scaling process starts by identifying the minimum and maximum values within the MFCC or Mel-Spectrogram matrix. The smallest value in the matrix is assigned to the first color in the colormap and the largest value is mapped to the last color. Intermediate values are linearly interpolated to assign corresponding colors between these two extremes. 
Converting MFCC and Mel-spectrogram features to images enables the use of pre-trained CNNs. In addition, this conversion allows the preservation of the time and frequency dimensions necessary for accurate audio analysis.
\subsection{Classification}
To assess how well classical machine learning models performed in detecting speech-related issues among children, we compared three classifiers: SVM, KNN, and Decision Trees. The speech signal of the child is first recorded as part of the diagnostic procedure. It is then put through a number of preprocessing stages to extract MFCC features which are subsequently supplied into the SVM, KNN, and Decision Tree classifiers. SVM, KNN, and Decision Trees are selected as baseline models for comparison. SVM, KNN, and Decision Trees have shown their effectiveness in speech recognition in \cite{smith2001speech} and \cite{sun2019decision} respectively. 

We also used ResNet18 deep neural networks in this work. ResNet, Residual Neural Network, is an advanced neural network architecture developed to solve the issue of vanishing gradients in deep neural networks by implementing skip connections - called residual connections - that bypass one or more layers. The residual block, in ResNet, consists of two convolutional layers with batch normalization and rectified linear unit (ReLU) activation functions, followed by a skip connection that adds the input to the output of the second convolutional layer. This skip connection preserves the original input information and helps propagate gradients more effectively during training. The inputs of the ResNet classifier, unlike SVM, KNN, and Decision Trees, are images obtained after the conversion of MFCC and Mel-Spectrogram matrices. To find out which model works best at diagnosing particular speech patterns or disorders, each classifier is assessed based on specific performance metrics. 

\section{Experiments and Results}
 All experiments were carried out using a Lenovo, i7 processor with 4 Cores and 8 Logical Processors, a RAM of 16GB.  We used Java to develop the game provided in the therapeutic module and MATLAB for the speech analysis used in the diagnostic module.  
\subsection{Datasets}
\begin{figure}[h] 
    \centering
    \begin{adjustbox}{max height=0.5\textheight, max width=1.5\textwidth} 
        \includegraphics{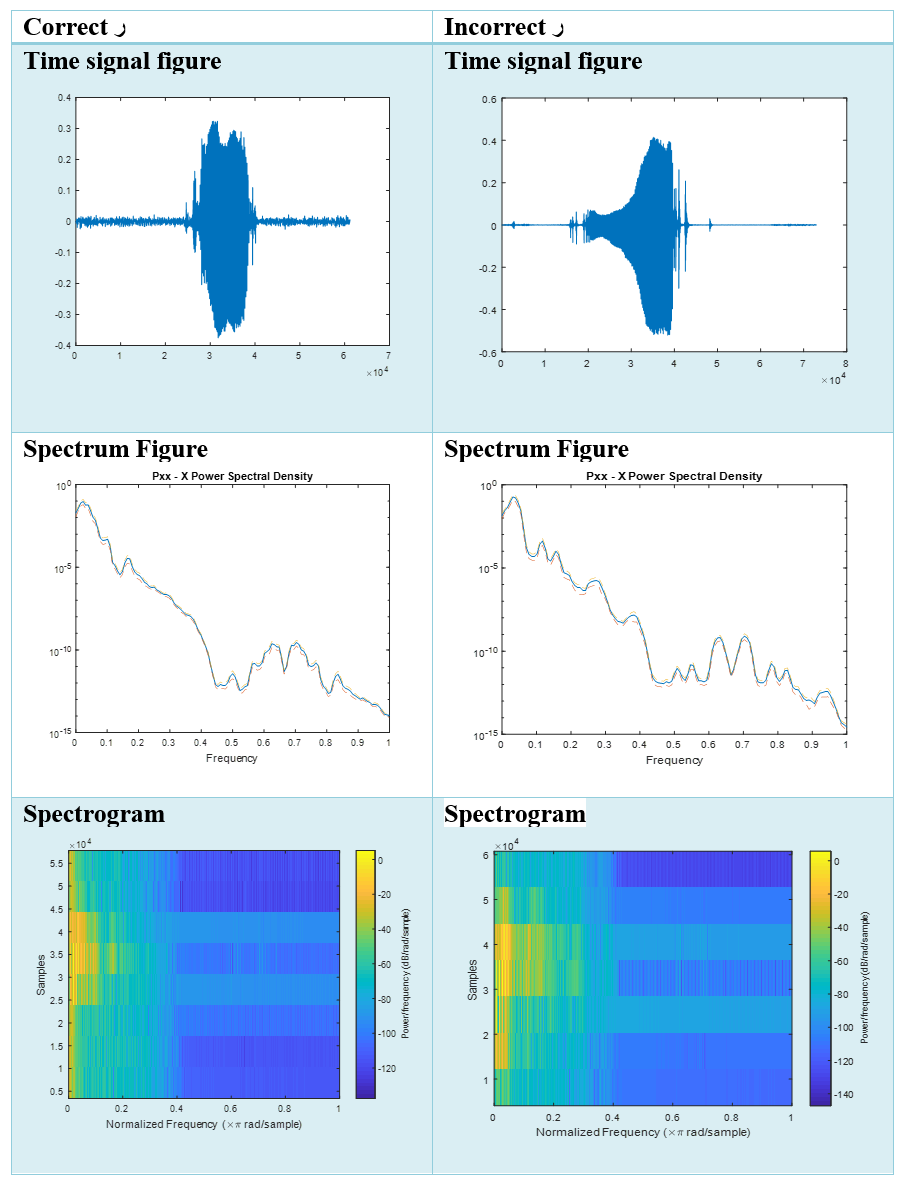}
    \end{adjustbox}
    \caption{Correct and incorrect Pronunciation of Letter "Raa"}
    \label{fig:Spectrum_Raa}
\end{figure}

In this work, we used two datasets. We collected the first dataset from a childcare center from kids aged between 5 and 7 years old. We recorded 10 correct and 10 incorrect letters. We focused on three letters "Raa", "Ghaa", and "Thaa".These letters have unique phonetic characteristics that can emphasize different aspects of speech processing. We recorded the kids’ voices and saved them as waveform (.wav) files. For data augmentation, we used a pitch variation factor to reduce the effect on the pronunciation. Afterwards, we augmented the dataset to 100 samples per letter. 
The second dataset is the Arabic dataset for alphabet pronunciation classification \cite{app11062508}. It is a balanced dataset that contains 8120 audio records for all Arabic letters providing a broader range of pronunciations and phonetic variations including both child and adult pronunciations. This allows us to capture a variety of pronunciation patterns, which can enhance the model's robustness.
The spectrum of correct and incorrect pronunciations of the letter "Raa" for both adults and kids are represented in Figure~\ref{fig:Spectrum_Raa}. 
\subsection{Performance Evaluation}
We compare the different classification techniques using the following performance criteria: accuracy, precision, recall, and F1-score. The accuracy (see Eq.\ref{eq:Accuracy}) measures the proportion of correctly classified pronounced letters among all audio instances. The recall (see Eq.\ref{eq:Recall}) measures the proportion of correctly predicted positive instances among all actual positive instances. The precision (see Eq.\ref{eq:Precision}) measures the proportion of correctly predicted positive instances among all instances predicted as positive. Finally, the F1-score (see Eq.\ref{eq:F1}) gives the harmonic mean of precision and recall.
\begin{equation}
\text{Accuracy} = \frac{TP + TN}{TP + FP + TN + FN}
\label{eq:Accuracy}
\end{equation}
\begin{equation}
\text{Recall} = \frac{TP}{TP + FN}
\label{eq:Recall}
\end{equation}
\begin{equation}
\text{Precision} = \frac{TP}{TP + FP}
\label{eq:Precision}
\end{equation}
\begin{equation}
\text{F1} = \frac{2 \cdot \text{Precision} \cdot \text{Recall}}{\text{Precision} + \text{Recall}}
\label{eq:F1}
\end{equation}

TP  (True Positives) and TN (True Negatives) represent the number of instances correctly predicted as positive and negative, respectively. FP (False Positives) and FN (False Negatives) represent the number of instances incorrectly predicted as positive or negative, respectively.
\subsection{Diagnostic Module Results}
\subsubsection{Classifications with classical machine learning:}
First, we used the MFCC features to compare SVM, KNN and Decision Tree classifier models. We used 10-fold cross-validation to divide the data into training and testing sets. In Table~\ref{tab:classifier_metrics}, we presented the results of applying the classical classifiers with for dataset 1  and dataset 2 after hyperparameters tuning using Bayesian optimization. 
\renewcommand{\arraystretch}{2} 
\vspace{-10pt} 
\begin{table}[H]
    \centering
    \begin{adjustbox}{max width=\textwidth}
        \begin{tabular}{|c|c|c|c|c|c|c|c|c|c|c|c|c|}
            \hline
            \textbf{}          & \multicolumn{4}{c|}{\textbf{SVM}} & \multicolumn{4}{c|}{\textbf{KNN}} & \multicolumn{4}{c|}{\textbf{Decision Tree}} \\ \hline
            \textbf{}          & \cellcolor[HTML]{FFFFFF}\textbf{Precision} & \cellcolor[HTML]{FFFFFF}\textbf{Recall} & \cellcolor[HTML]{FFFFFF}\textbf{F1-Score} & \textbf{Accuracy} & \cellcolor[HTML]{FFFFFF}\textbf{Precision} & \cellcolor[HTML]{FFFFFF}\textbf{Recall} & \cellcolor[HTML]{FFFFFF}\textbf{F1-Score} & \textbf{Accuracy} & \cellcolor[HTML]{FFFFFF}\textbf{Precision} & \cellcolor[HTML]{FFFFFF}\textbf{Recall} & \cellcolor[HTML]{FFFFFF}\textbf{F1-Score} & \textbf{Accuracy} \\ \hline
            \textbf{Dataset 1} & 89.74\% & 83.33\% & 86.42\% & 86.42\% & 95.35\% & 97.62\% & 96.47\% & 96.3\% & 82.54\% & 85.25\% & 83.87\% & 83.61\% \\ \hline
            \textbf{Dataset 2} & 92.2\% & 91.75\% & 91.98\% & 92\% & 93.9\% & 94.84\% & 94.36\% & 94.33\% & 81.95\% & 80.94\% & 81.44\% & 81.57\% \\ \hline
        \end{tabular}
    \end{adjustbox}
    \vspace{5pt} 
    \caption{Performance Results Obtained by Applying Different Classical Classifiers Using Dataset 1 and 2}
    \label{tab:classifier_metrics}
\end{table}
\vspace{-30pt} 
As shown in Table~\ref{tab:classifier_metrics}, KNN gives the best results for dataset 1 and dataset 2. This shows the robustness of KNN in differentiating between correct and incorrect pronunciation even for slight variations in features. Also, we notice that SVM is less effective in the case of a small dataset as it may not capture the exact structure of the data.
\vspace{-10pt}
\subsubsection{Classification with ResNet18 Applied on MFCC Images:} 
\begin{figure}[H]
     \centering
     \begin{subfigure}[b]{0.3\textwidth}
         \centering
         \includegraphics[width=0.75\textwidth]{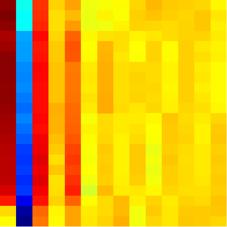}
         \caption{MFCC Image}
         \label{fig:MFCC_Image}
     \end{subfigure}
     \hspace{0.5cm}
     \begin{subfigure}[b]{0.3\textwidth}
         \centering
         \includegraphics[width=0.75\textwidth]{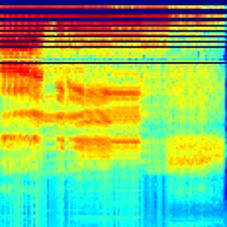}
         \caption{Spectrogram Image}
         \label{fig:Spectrogram Image}
     \end{subfigure}
           \caption{Spectrogram and MFCC Images}
        \label{fig:Spectrogram and MFCC Images}
\end{figure}

\begin{figure}[h]
     \centering
     \begin{subfigure}[b]{0.45\textwidth}
         \centering
         \includegraphics[width=\textwidth]{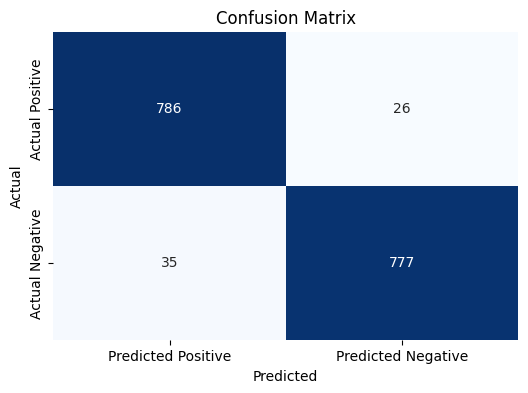}
         \caption{Confusion Matrix of the Classification of MFCC Images Using ResNet18}
         \label{fig:Confusion Matrix of the classification of MFCC images using ResNet18}
     \end{subfigure}
     \hfill
     \begin{subfigure}[b]{0.45\textwidth}
         \centering
         \includegraphics[width=\textwidth]{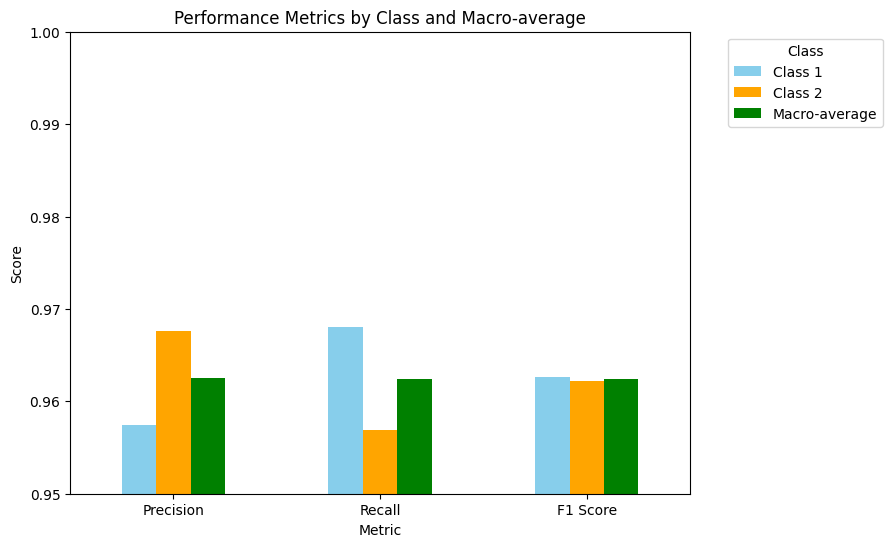}
         \caption{Precision, Recall, and F1-score Results of Obtained by Classifying MFCC Images Using ResNet18}
         \label{fig:Spectrogram Image}
     \end{subfigure}
           \caption{Classification Results of MFCC images Using ResNet18}
        \label{fig:Results_ResNet18}
\end{figure}

After extracting the MFCC coefficients from each speech signal, each MFCC matrix is converted into a color image. An example of an MFCC image is shown in Figure~\ref{fig:MFCC_Image}. We divide the dataset randomly into two sets, i.e., training and testing sets. The training set contains 80\% of the whole dataset. The testing set contains the remaining 20\% of the dataset, considering that it only contains novel data not dealt with in the training process. We tested ResNet18 with different training parameters and after an empirical analysis, we fine-tuned our model with a minibatch size of 100, 30 epochs, a 0.0001 learning rate, and an Adam optimizer. The validation accuracy of the ResNet model reached 97.53\% for dataset 1 and 96.244\% for dataset 2. In Table~\ref{tab:ResNet18_Results_MFCC}, we show the results of ResNet18 with MFCC images for both dataset 1 and dataset 2. Also, the confusion matrix and the performance results per class for dataset 2 are given in Figure~\ref{fig:Results_ResNet18}. We notice that our model has high precision, recall, and F1-score for both classes. This shows its robustness in identifying both positive and negative instances.
\vspace{-30pt}
\begin{table}[H]
    \centering
    \begin{adjustbox}{max width=\textwidth}
        \begin{tabular}{|l|c|c|c|c|}
            \hline
            \textbf{}          & \cellcolor[HTML]{FFFFFF}\textbf{Precision} & \cellcolor[HTML]{FFFFFF}\textbf{Recall} & \cellcolor[HTML]{FFFFFF}\textbf{F1-Score} & \textbf{Accuracy} \\ \hline
            \textbf{Dataset 1} &  97.63\%                                          &                             97.73\%            &                                    97.62\%       &    97.53\%               \\ \hline
            \textbf{Dataset 2} & 96.25\%                                    & 96.24\%                                 & 96.24\%                                   & 96.24\%           \\ \hline
        \end{tabular}
    \end{adjustbox}
    \vspace{5pt} 
    \caption{ResNet18 Results Obtained by Using Dataset 1 and 2 and MFCC Images}
    \label{tab:ResNet18_Results_MFCC}
\end{table}
\vspace{-30pt}
\vspace{-10pt}
\subsubsection{Classifications with ResNet18 Applied on Mel-Spectrogram Images}
\begin{figure}[h]
     \centering
     \begin{subfigure}[b]{0.45\textwidth}
         \centering
         \includegraphics[width=\textwidth]{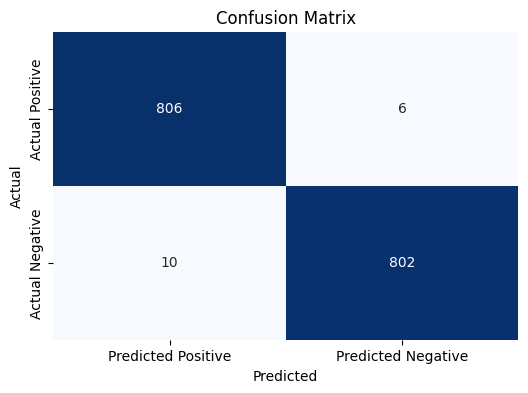}
         \caption{Confusion Matrix of the Classification of Mel-Spectrogram Images Using ResNet18}
         \label{fig:Confusion Matrix of the classification of Melspectrogram images using ResNet18}
     \end{subfigure}
     \hfill
     \begin{subfigure}[b]{0.45\textwidth}
         \centering
         \includegraphics[width=\textwidth]{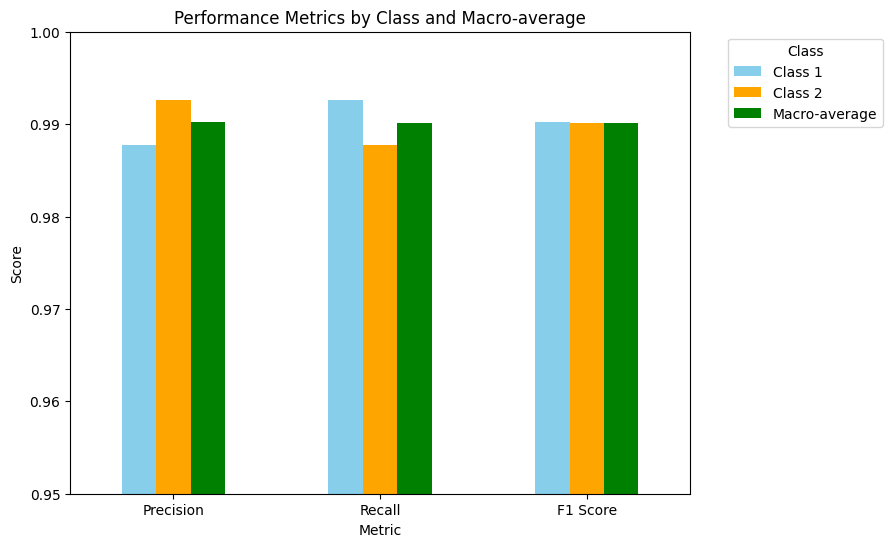}
         \caption{Precision, Recall, and F1-score Results of Obtained by Classifying Mel-Spectrogram Images Using ResNet18}
         \label{fig:Precision, Recall, and F1-score of the classification of Mel-Spectrogram images using ResNet18}
     \end{subfigure}
           \caption{Classification Results of Mel-Spectrogram images Using ResNet18}
        \label{fig:Results_ResNet18_MelSpectrogram}
\end{figure}

First, we converted the Mel-Spectrogram matrices to color images. An example of a Mel-Spectrogram image is shown in Figure~\ref{fig:Spectrogram Image}. The audio signal images of the training set are used to train the ResNet 18. We tested ResNet18 with different training parameters and after an empirical analysis, we fine-tuned our model with a minibatch size of 100, 30 epochs, a 0.0001 learning rate, and an Adam optimizer. The validation accuracy of the ResNet model reached an accuracy of 98.77\% for dataset 1 and  99.015\% for dataset 2. In Table~\ref{tab:ResNet18_Results_MelSpectrogram}, we provide the results of ResNet18 with Mel-Spectrogram images for both dataset 1 and dataset 2. In addition, the confusion matrix and the performance metrics per class for dataset 2 are given in Figure~\ref{fig:Results_ResNet18_MelSpectrogram}. We notice that our model has high precision, recall, and F1-score for both classes. This shows its robustness in identifying both positive and negative instances.
\vspace{-15pt}
\begin{table}[h]
    \centering
    \begin{adjustbox}{max width=\textwidth}
        \begin{tabular}{|l|c|c|c|c|}
            \hline
            \textbf{}          & \cellcolor[HTML]{FFFFFF}\textbf{Precision} & \cellcolor[HTML]{FFFFFF}\textbf{Recall} & \cellcolor[HTML]{FFFFFF}\textbf{F1-Score} & \textbf{Accuracy} \\ \hline
            \textbf{Dataset 1} &  98.72\%                                          &                   98.75\%                      &  98.73\%                                         &       98.77\%            \\ \hline
            \textbf{Dataset 2} & 99.01\%                                    & 99.01\%                                 & 99.01\%                                   & 99.015\%           \\ \hline
        \end{tabular}
    \end{adjustbox}
    \vspace{5pt} 
    \caption{ResNet18 Results Obtained by Using Dataset 1 and 2 and Mel-Spectrogram images}
    \label{tab:ResNet18_Results_MelSpectrogram}
\end{table}
\vspace{-35pt}

We notice that the use of ResNet18 with Mel-Spectrogram images outperforms ResNet18 with MFCC. We interpret this result by the fact that Mel-Spectrogram provides a detailed 2D time-frequency representation of the audio signal whereas MFCC is a compressed version of the Mel-Spectrogram.
\subsection{Therapeutic module}
\begin{figure}[h]
     \centering
     \begin{subfigure}[b]{0.45\textwidth}
         \centering
         \includegraphics[width=\textwidth]{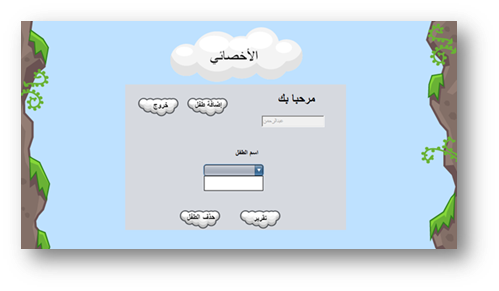}
         \caption{Therapist Interface}
         \label{fig:Doctor_interface}
     \end{subfigure}
     \hfill
     \begin{subfigure}[b]{0.45\textwidth}
         \centering
         \includegraphics[width=\textwidth]{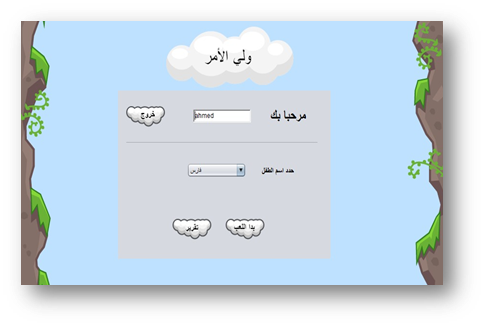}
         \caption{Parent Interface}
         \label{fig:Parent_Interface}
     \end{subfigure}
           \caption{Therapeutic Module Interfaces}
        \label{fig:Therapeutic_module_interfaces}
\end{figure}

\begin{figure}[h]
     \centering
           
        \includegraphics[width=0.75\textwidth]{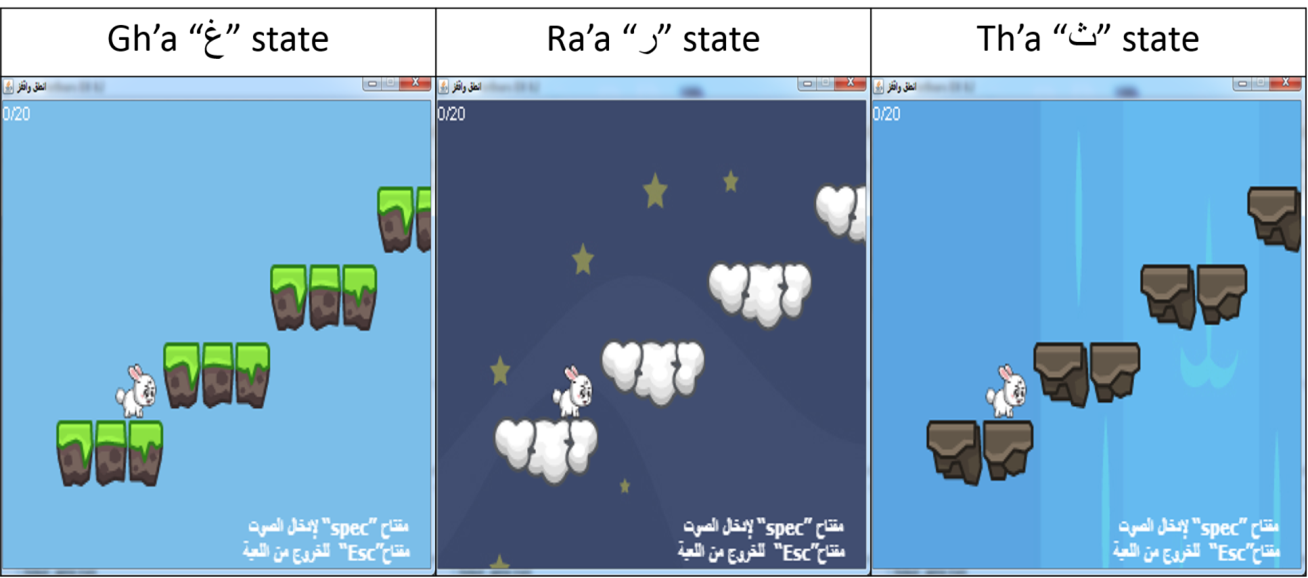}
         \caption{The game}
         \label{fig:The_game}
 \end{figure}

In this module, \textbf{ArPA} offers a game that starts with a registration interface for the user as a parent or therapist. After that, the kid's information such as age and gender are saved. After the registration phase, the child can use the game under the parent's or the therapist's control, see Figure~\ref{fig:Therapeutic_module_interfaces}. The game interacts with the kid via the voice entry only. In case, the kid's pronunciation of the letter is correct, a rabbit jumps to the next level otherwise it remains on the same level, see Figure~\ref{fig:The_game}. The history of the levels reached by the kid for each letter is saved in a report that could be viewed later. A report could be generated and used by the therapist to analyze the kid's amelioration.

\section{Conclusion}
This paper introduces \textbf{ArPA}, an innovative application developed for Arabic-speaking children with speech problems, focusing on their distinct speech characteristics rather than targeting adults or non-native speakers. ArPA includes diagnostic and therapeutic components that use sophisticated deep learning classifiers, such as ResNet18, to effectively identify mispronunciations, showcasing high precision, recall, and F1 scores in testing outcomes. 
However, we emphasize that the main issue encountered during the development of this paper was the limited dataset size. Subsequent research will focus on increasing the considered dataset with additional samples of children's speech, improving gamified therapy strategies for increased participation, incorporating more deep learning structures and transfer learning techniques such as audio deep learning, and carrying out long-term assessments to evaluate the effects of \textbf{ArPA} on children's language development and pronunciation abilities.

%
%

\bibliographystyle{splncs04}  
\bibliography{references}  
\end{document}